# Simultaneous nanoscale excitation and emission mapping by cathodoluminescence

**AUTHOR NAMES**

*Taeko Matsukata[1], Shintaro Ogura[1], F. Javier García de Abajo[2,3], Takumi Sannomiya[1]*

**AUTHOR ADDRESS**

[1] Department of Materials Science and Technology, Tokyo Institute of Technology, 4259 Nagatsuta Midoriku, Yokohama 226-8503, Japan

[2] ICFO-Institut de Ciencies Fotoniques, The Barcelona Institute of Science and Technology, 08860 Castelldefels (Barcelona), Spain

[3] ICREA-Institució Catalana de Recerca i Estudis Avancats, Passeig Lluís Companys 23, 08010 Barcelona, Spain

*Abstract*

Free-electron-based spectroscopies can reveal the nanoscale optical properties of semiconductor materials and nanophotonic devices with a spatial resolution far beyond the diffraction limit of light. However, the retrieved spatial information is constrained to the excitation space defined by the electron beam position, while information on the delocalization associated with the spatial extension of the probed optical modes in the specimen has so far been missing, despite its relevance in ruling the optical properties of nanostructures. In this study, we demonstrate a cathodoluminescence method that can access both excitation and emission spaces at the nanoscale, illustrating the power of such simultaneous excitation and emission mapping technique by revealing a sub-wavelength emission position modulation as well as by visualizing electromagnetic energy transport in nanoplasmonic systems. Besides the fundamental interest of these results, our technique grants us access into previously inaccessible nanoscale optical properties.

Over the last decades, nanoscopic optical characterization using free electron beams (e-beams) has attracted considerable attention in various research fields, such as nanophotonics and materials science, while fundamental physics benefits as well from the ability of e-beams to access optical states beyond the diffraction limit of light.[1-3] Optical measurement techniques based on free electrons are classified into electron energy-loss spectroscopy (EELS)[4, 5] and cathodoluminescence (CL),[6, 7] which are typically performed utilizing a scanning transmission electron microscope (STEM) or a scanning electron microscope (SEM) to obtain spectrally and spatially resolved optical information with a resolution down to a few nanometers or less.[8] Optical information in EELS measurements is extracted by recording the energy loss experienced by the incident electrons, while photons generated upon e-beam excitation are collected and analyzed in CL measurements. The nominal high spatial resolution in both approaches arises from the small size of the e-beam at the position of the specimen and the precision in the excitation position, although the optical properties of the sampled nanostructures are oftentimes dominated by relatively delocalized emission or scattering processes involving multiple optical modes and their coupling to radiation, rather than by the actual excitation. In particular, a most interesting piece of information in CL lies is the processes taking place in between excitation and emission, producing spatiotemporal energy flow across different modes of the specimen. Indeed, the excitation and emission positions do not match in general in optical systems such as semiconductors due to carrier diffusion, topologically protected edge-states capable of moving before out-coupling to radiation, cavities combined with localized emitters, and many others.[6, 9, 10] Although the CL approach is advantageous with respect to EELS for the characterization of emission, it does not directly reveal the position from which photons are emerging. This leaves us with a longstanding problem in the analysis of semiconductor optical devices, where excited carriers or excitons recombine and emit photons after spatial diffusion that reduces the effective spatial resolution in the CL measurement.[6, 11] Also, in nanophotonic antennas or waveguides utilizing coherent processes, determination of the emission position is important to analyze and engineer the electromagnetic energy transport and conversion.[12] To monitor the carrier diffusion in semiconductors, a method that combines near-field probing and electron beam excitation has been proposed,[13-15] which nevertheless suffers limitations in detection angle or polarization selectivity due to the near-field probing nature as well as the disturbance of the environment by the probe itself.

In this study, we demonstrate a simultaneous excitation and emission mapping (SEEM) approach based on CL and using a collimation parabolic mirror as an imaging element to project the three-dimensional (3D) information associated with the emission position on an arbitrarily chosen two-dimensional (2D) plane (Fig. 1a). To select the projection plane of the emission space, angle-resolved CL detection is employed. The current method extends in a nontrivial way previous CL capability, such as the evaluation of dispersion relations as well as the determination of emission directionality and interference. [7, 16-19] By synchronizing emission imaging with e-beam scanning, CL photon maps of a given spot in the emission image can be extracted. Since the resolution of the emission image is restricted by the diffraction limit of light, we adopt a spot analysis method borrowed from super-resolution microscopy (such as in photoactivated localization microscopy (PALM)), allowing us to quantitatively evaluate the spatial shift of emission spots in the emission images beyond the diffraction limit.[20] We apply the SEEM method to the analysis of different nanophotonic systems, and in particular, a silver nanowire in which subtle shifts of the emission spot are observed to be modulated by the presence of plasmon modes, as well as silver nanoholes in which we image multiple emission positions that reveal the involvement of electromagnetic energy transport.

**Emission imaging in STEM-CL**

The CL-emission imaging system is built in a STEM-CL setup equipped with an angle-resolved spectrum mapping system,[17, 18, 21] where a parabolic mirror is inserted at the sample position of the STEM instrument to collimate the light radiated from the sample (Fig.1). The emitted light, transferred to the optical system through a polarizer and an angle selection mask, is subsequently focused on a camera for emission imaging (Fig.1b). Simultaneous spectral measurement is also available by means of a beam splitter (see Supplementary Information (SI) for details on the setup) The collected CL signal is given by the overlap between the excitation region and the optical collection spot, which is back-projected from the detection optics (schematically illustrated in Fig. S1 in SI). Since each pixel on the camera can be optically traced onto a back-projected spot on the sample plane, the emission image on the camera corresponds to the optical spot scan on the specimen. The "excitation region" extends beyond the e-beam position due to the finite extension of the modes sampled in the specimen (i.e., wave propagation or diffusion). We use two different modes of operation, namely: (1) "emission imaging" corresponding to scanning the optical spot over the specimen for a fixed selected e-beam spot position (see Fig. 1c); and (2) "emission-spot-decomposed CL mapping", in which the e-beam is scanned while the optical spot is fixed at a selected position (see Fig. 1d).

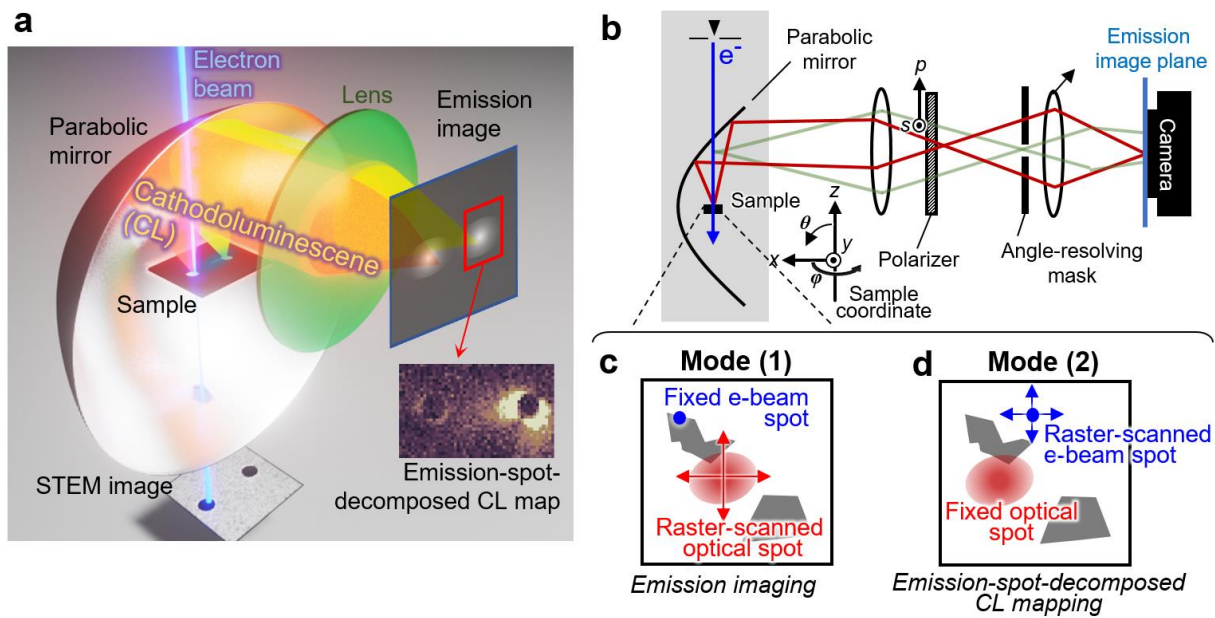

**Figure 1. Simultaneous excitation and emission mapping (SEEM) by cathodoluminescence.** (a) Illustration of emission imaging synchronized with the electron beam (e-beam) scanning. (b) Diagram of the optical path used to image the emission position spot. Ray paths associated with emission-space and reciprocal-angular-space imaging are drawn as red and green lines, respectively. The emission angle can be selected in the projected angular space with an angle-resolving pinhole mask, as shown in the diagram. (c,d) Conceptual illustration of signal acquisition in the two modes of operation introduced in this work: (c) emission imaging, in which the e-beam is fixed at a selected specimen position, while the optical collection spot is scanned, so that each image pixel corresponds to an optical spot position; and (d) emission-spot-decomposed CL mapping, in which the optical collection spot is fixed and the e-beam position is raster scanned. As a variant of mode (1), we also produce images integrated for a set of e-beam positions covering the entire specimen.

Using the parabolic mirror as the image forming element, the optical axis for the image formation can be selected by the angle-resolving mask placed on the angular space plane. Without the angle-resolving mask, the optical axis corresponds to the $x$ axis with a detection solid angle of around $3\pi$ steradians, and the imaging plane of the camera corresponds to the $y$-$z$ plane of the sample space (see SI for these imaging conditions). Under such conditions, a sample shift along the $x$ direction produces a defocus of the emission image. To obtain an emission image corresponding to a $x$-$y$-plane projection (i.e., the STEM imaging plane), the optical axis should be along the $z$ axis, implying that only radiation emitted along the upward direction should be collected. Figure 2 shows an emission image demonstration (mode (1), see Fig. 1c) with upward detection ($\theta = 0°$), revealing a shift in the emission spot when physically displacing the sample (a ~2 µm ZnS:Pb particle), as schematically illustrated in panels a, d, and g. In this measurement, a single emission spot image is acquired by integrating the signal while scanning the e-beam over the entire area of the STEM image. Thus, the imaged emission spot corresponds in fact to the image of the entire particle. As the specimen is moved along the $x$ and $y$ axis by ±2 µm, the emission spot shifts accordingly in the emission images (Fig. 2c and f), faithfully mimicking the sample shift in the STEM images (Fig. 2b and e). With a $z$ specimen shift, which corresponds to defocus in the STEM image (Fig. 2h), the emission image in Fig. 2i is also defocused. These sets of results show that the projection plane of the three-dimensional emission position space can be arbitrarily chosen by selecting the emission angle as, for example, the $x$-$y$ or $y$-$z$ plane (see also the SI).

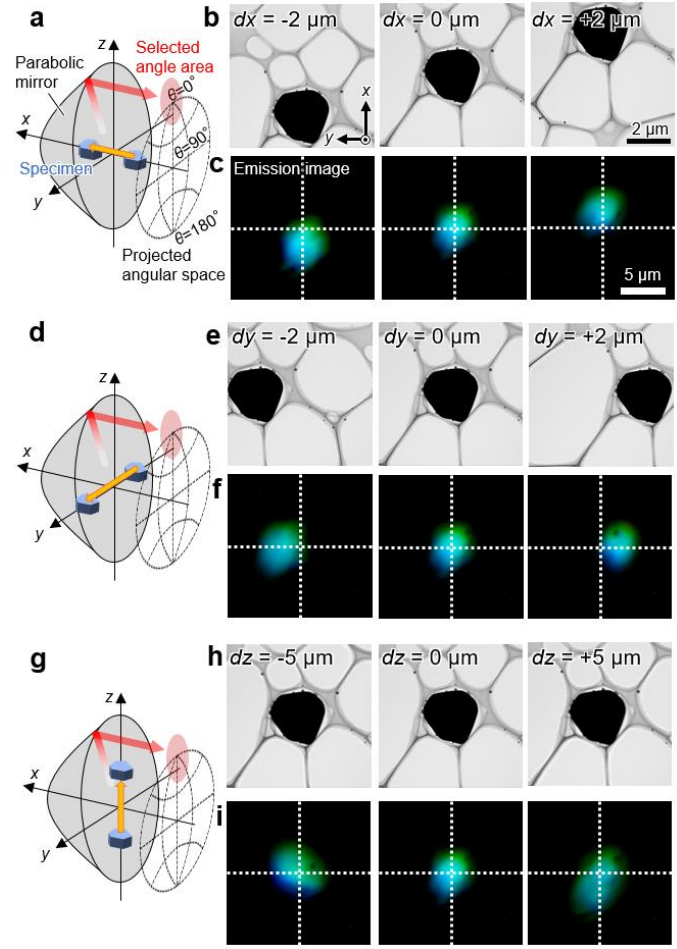

**Figure 2. Emission spot shift produced upon sample position displacement.** (a, d, g) Schematic illustrations of sample position displacements along (a) *x*, (d) *y*, and (g) *z* directions for a probed ZnS:Pb particle. The angle-resolving pinhole mask is inserted to select upward radiation. (b, e, h) STEM bright-field images of the probed particle. The sample is moved by ± 2 µm along the *x* and *y* directions, and by ± 5 µm along *z*. (c, f, i) Emission images (see Fig. 1c) obtained by integrating the e-beam region over the whole STEM images. The angle-resolving mask transmits light around $(\theta, \varphi) = (0°, 0°)$ within a detection solid angle of 0.2 sr.

**Emission-spot-decomposed CL mapping**

Since an emission image (as a function of the position of the optical spot) is obtained for each e-beam position, a four-dimensional (4D) data set is obtained by synchronizing the emission image acquisition with a raster scan of the e-beam. The data representation of the two operation modes (Fig. 1c and 1d) is possible by post-processing the 4D data set (see SI for details). In this section, we present an example of mode (2) (see Fig.1d). The results obtained by applying mode (1) (see Fig. 1c) to the same dataset are shown in SI. The angle-resolving mask is inserted such that the imaging plane corresponds to the *x-y* sample plane by selecting the upward emission from ZnS:Pb particles, as illustrated in Fig. 3a. Figure 3c shows an integrated emission image over the entire e-beam scan area (Fig. 3b), showing three spatially separated emission spots in the integrated emission image corresponding to the three particles in the STEM bright-field image (Fig.3c). Figure 3d shows emission-spot-decomposed CL maps constructed by integrating over e-beam excitation positions around selected particles. The signal integration area of the excitation position is indicated by the dashed rectangles (I)-(III) in Fig. 3c. The emission-spot-decomposed CL map integrating all the three emission spots in Fig. 3d-(I) shows clear particle shapes that are in good correspondence with the STEM bright-field image. Figure 3d-(II) shows the emission-spot-decomposed CL map integrated over the upper right emission spot, as indicated by the rectangular region (II) in Fig. 3c. Only the upper right particle is imaged in the CL map corresponding to the selected emission spot. Similarly, as shown in Fig. 3d-(III), the emission-spot-decomposed CL map corresponding to the region (III) containing the lower spot in Fig. 3c shows the CL particle image of only the lower particle. Thus, the emission-spot-decomposed CL mapping can directly visualize the correlation between the excitation and emission positions. For this particular instance of separated particles, the correlation is complete, in contrast to connected regions in nanowires and perforated films, in which the correlation between excitation and emission reveals information about intermediate processes, as we discuss below. As a reverse analysis, the emission image of a certain excitation can also be extracted from the 4D data set of the SEEM measurement (see details in SI).

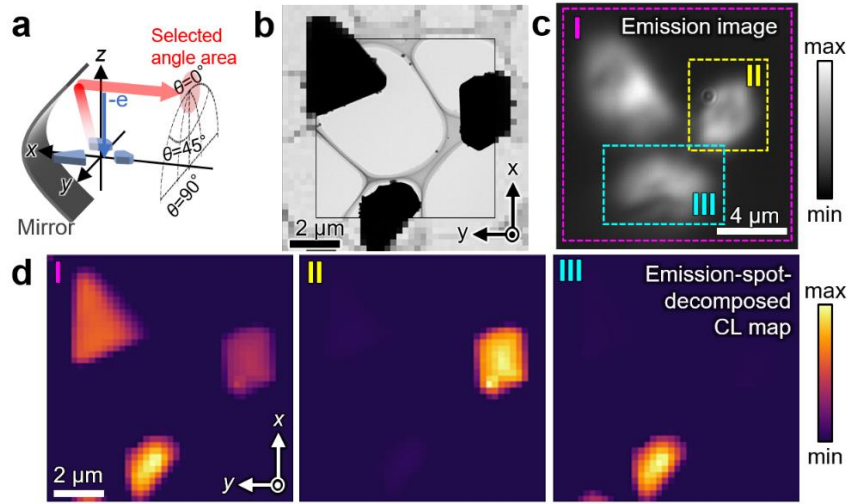

**Figure 3. Emission-spot-decomposed CL mapping:** (a) Schematic illustration of the measurement conditions for a selection of emission angles $(\theta, \varphi) = (0°, 0°)$ with a detection solid angle range of 0.2 steradians such that the *x-y* emission position plane is imaged. (b) STEM bright-field image of the probed ZnS:Pb particles. A fine scan square image is inserted. (c) Emission image (mode 1, see Fig. 1c) with the signal integrated over the whole range of e-beam excitation positions. (d) Emission-spot-decomposed CL maps (see Fig. 1d) integrated over the emission image within the rectangular regions I–III indicated in panel (c). No wavelength filtering is performed.

**Beating the diffraction limit in the evaluation of the emission spot shift**

Although the spatial resolution to separate two emission spots is limited by Rayleigh's criterium, the position of a single spot can be determined well below the diffraction limit of light from the location of the intensity maximum, in analogy to optical super-resolution techniques such as PALM.[20] Since the actual profile of the point spread function (PSF) of the emission spot slightly changes with position due to the parabolic mirror optics, we adopt a center-of-mass (COM) evaluation to determine the emission spot position instead of the PSF fitting used in the PALM method. To examine this evaluation method, we measure transition radiation from a planar metal layer.[22] Since transition radiation can be attributed to a perpendicular dipole excited exactly at the position of the e-beam, the emission position can be precisely controlled with the resolution of the electron microscope. To efficiently collect the transition radiation, we insert the angle-resolving pinhole mask at a detection angle of $\theta = 45°$ and set the solid angle range to 1.8 sr, as schematically shown by the red hatched area in the angular space in Fig. 4a. Since emitter shifts along the $x$ and $z$ directions are both projected on the vertical axis in the emission position image under this observation conditions (see Fig. S4 in SI), we here evaluate only the emission spot shift in the horizontal direction corresponding to the $y$ direction in the sample space. Figure 4b shows the emission image of a representative excitation position, where the intensity spreads approximately over 500 nm due to light diffraction. The calculated COM position is marked as a blue dot. For three representative horizontal scans, the COM position of each excitation point is plotted on the emission image integrated over the three scans, as shown in Fig. 4c. To summarize the evaluation, we plot the emission spot shift in Fig. 4d for an e-beam scan area of 1.6 μm × 1.6 μm as a function of the excitation position shift (in steps of 80 nm in $x$ position). The distance between two emission spots is evaluated for all possible combinations in each horizontal scanline. The horizontal axis in Fig. 4d corresponds to the shift in the excitation position by the e-beam and the vertical axis stands for the average $y$-shift of the emission spot as resolved from the COM. The error (standard deviation of the measurement) increases as the excitation position shift becomes larger, which indicates that the shape of the spot deforms for large position shifts (see Fig. S5 in SI). For a shift in excitation position of up to 1 μm, the error in the evaluation of the emission position is ≤50 nm, well below the diffraction limit of light. For smaller emission position shifts, the accuracy can reach down to ~10 nm, as shown in Fig. 4e. We thus conclude that super-resolution beyond the light diffraction limit can be achieved in SEEM on the determination of the CL emission position.

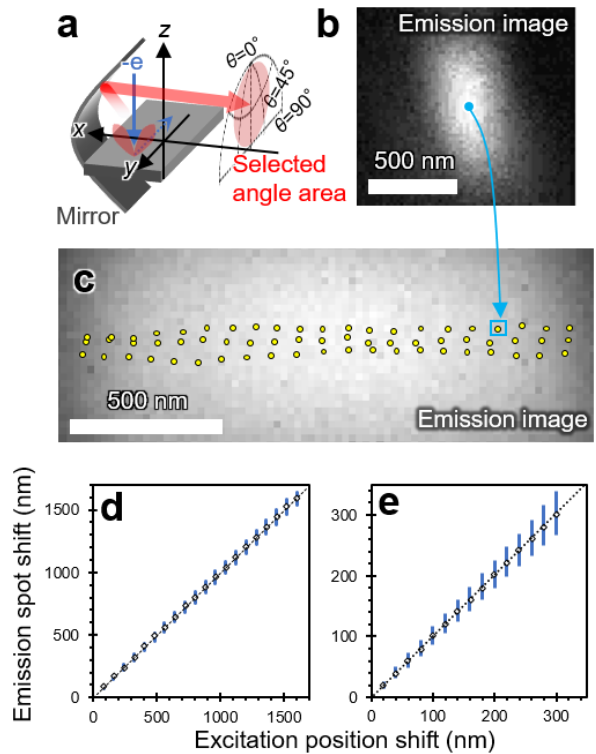

**Figure 4. Evaluation of the emission spot position of transition radiation from a silver substrate.** (a) Schematic illustration of transition radiation measurement from a silver film. The pinhole mask area is represented as a red circle in the emission-angle space. (b) Representative

emission spot image (using the method sketched in Fig.1c) for a selected e-beam position, with the center-of-mass (COM) position of the emission indicated by a blue dot. The emission spot image corresponds to the e-beam position indicated by blue rectangle in panel c. (c) Emission image superimposed over three different horizontal scans. The yellow points show the evaluated COM position of each spot image as the excitation position of the e-beam is scanned with a step of 80 nm. (d,e) Summary plots of the emission position shift resolved from the COM of each emission spot while shifting the e-beam position. The error bar corresponds to the standard deviation. The dashed lines show the ideal proportionality.

### Shift of the emission in a plasmonic nanowire

We now evaluate slight shifts of the emission spot in the presence of optical modes in a plasmonic nanowire as a result of the delocalized nature of such modes. When a fast electron passes through or near a metal nanowire, an electric dipole is excited at the surface near the e-beam spot, giving rise to a localized emission source, in a similar manner as TR on a flat metal surface. Simultaneously with this localized dipole, surface plasmon modes of the wire are also excited (in response to such dipole), resulting in nodes/antinodes along the wire axis[23-25], which can perturb the resulting optical emission, which is no longer concentrated at the e-beam position. We explore this effect by examining a 1.5 µm long silver nanowire of 95 nm in diameter with its major axis oriented parallel to the $y$ direction (see Fig. 5a and b). As illustrated in Fig. 5a, by using the angle-resolving mask and a polarizer, only $p$ polarized emission around $\theta = 0°$ with a detection solid angle of 1.14 steradian is acquired to selectively observe the local polarization along the short axis of the wire parallel to the support film (a 20 nm thick free-standing elastic carbon membrane). We choose the wavelength range as 550-700 nm, in which the wire plasmon mode is most clearly observed. The mask area is schematically illustrated as a red circle in Fig.5a. In a conventional CL map, as shown in Fig. 5c, a high-intensity contrast with multiple antinodes is visible along the upper and lower edges of the wire, indicating the presence of a wire mode.[23-25] The high-intensity hotspot location appears alternatively at the upper and lower edges along the wire axis. This zig-zag contrast is due to the local polarization across the short axis interfering with the charge distribution of the wire mode as well as to the detection asymmetry originating from the circular aperture inserted around $\theta = 0°$, which covers a larger solid angle in the $x$-positive direction.[17] Indeed, a boundary-element method (BEM) simulation (Fig. 5g) for the same wire geometry corroborates the emergence of a clear standing wave pattern along the wire when no angle or polarization selection is applied. The zig-zag hotspot distribution is also reproduced with the polarization- and angle-selection under the experimental conditions, as shown in Fig. 5h (see details of the BEM calculations in Methods).

To examine how this interplay between the local dipole and the plasmon mode influences the emission position, we evaluate the shift of the emission spot position with respect to the excitation e-beam position. Three excitation spots indicated in Fig. 5b are chosen as representative positions and the corresponding three emission images are collected, as shown in Fig. 5d as a superposition with different colors. The emission spot seems to follow the change in the excitation position, indicating that localized emission from the electron beam position is dominant. The emission-spot-decomposed CL map using the corresponding emission position locations also supports the primary contribution of localized emission at the e-beam position (Fig. 5e, operation mode (2), see Fig. 1d). To more quantitatively corroborate the correlation between the shifts in the emission spot and e-beam excitation position, we perform a COM analysis of the emission spot similar to the previous discussion in Fig.4. In Fig 5f, the COM position is plotted as a function of the excitation position along the $y$ axis (long axis of the wire). The COM position plot reveals modulated features, resulting in deviations from a linear relation. This modulation period roughly corresponds to the antinode period, indicating some perturbation related to the standing waves associated with plasmon modes propagating along the wise. The simulated COM analysis shown in Fig. 5i further confirms this modulation of the spot position. The antinode positions and the corresponding spot shift positions are indicated by the dashed lines in Fig. 5c and f for the experiment and in Fig. 5h and i for the simulation. The spot position tends to be shifted towards the antinode positions the wire mode. Thus, using the SEEM method, we reveal that the CL emission position from the plasmonic nanowire excited by an e-beam is perturbated by the presence of plasmon modes and that the emission position tends to be attracted towards the antinode positions due to field propagation associated with those plasmons.

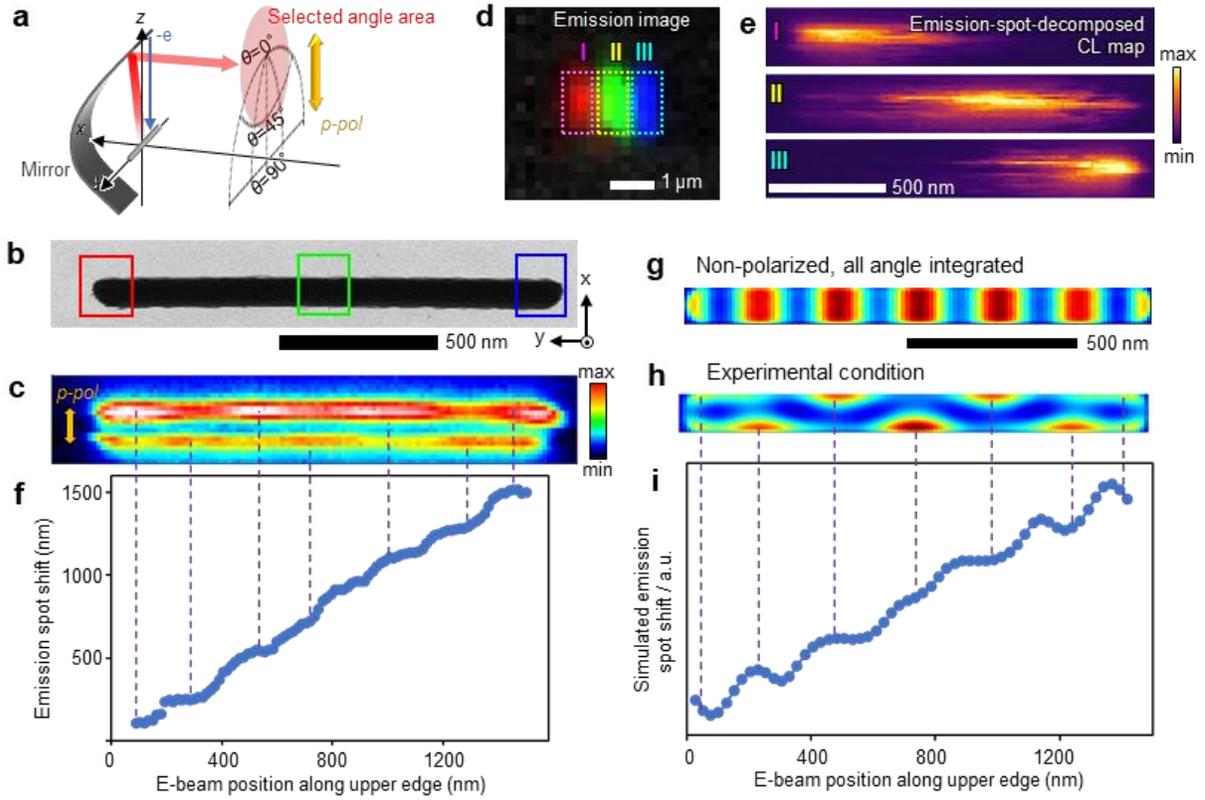

**Figure 5. Emission imaging of a silver nanowire.** (a) Schematic illustration of the measurement. The pinhole mask is inserted to select upward radiation with *p* polarization at a detection angle ($\theta$, $\varphi$) = (0°, 0°) with a solid angle of 1.14 sr, which is indicated by the red area in the emission angle space. The wavelength range is chosen as 550-700 nm. (b) STEM bright-field image of the silver nanowire (1.5 µm length and 95 nm diameter) and (c) corresponding CL map. The direction of *p* polarization is schematically shown by the yellow arrow in panels a and c. (d) Superimposed emission images (operation mode (1), see Fig. 1c) with the emission signals integrated over e-beam positions in the left edge (red), center (green), and right edge (blue) of the wire (see color-coordinated rectangles indicating the e-beam position regions in panel (b)). (e) Emission-spot-decomposed CL photon maps (operation mode (2), see Fig. 1d) with the emission signals integrated over the emission image in the areas (I)–(III) in panel d. (f) Relationship between the measured shifts in excitation position and emission COM from the left edge in the *y* (horizontal)

direction. The emission spot at each excitation position along *y* is evaluated along the upper edge of the wire in the *x* direction. (g) Simulated unpolarized CL map with the signal integrated over all angles. (h) Angle- and polarization-selected CL map under the same conditions as in the experiment. (i) Simulated COM shift as a function of e-beam position. The e-beam is scanned along the upper edge of the wire, as in the experiment of panel f. The wavelength for the simulation is 631.8 nm.

**Emission from metal holes coupled to surface-plasmon polaritons**

The second model system with (in principle) non-leaky wave propagation is a metal thin film perforated with holes, which supports both propagating surface-plasmon polaritons (SPPs) and localized surface-plasmon modes at the holes.[26, 27] The local modes can be excited by the e-beam directly hitting the hole edge as well as through scattering of SPPs induced by the e-beam on a flat silver area distant from the hole location.[16, 28] Since the e-beam incident on a flat silver surface generates transition radiation together with SPPs, two spatially separated emitters corresponding to the e-beam position and the hole are present. We use a slit-type mask with *p* polarization to detect both the transition radiation and the local mode, as illustrated in Fig. 6a. The diameter of the observed silver hole is 260 nm, as shown in the STEM dark-field image in Fig. 6b. Figure 6c shows the emission images integrated over the e-beam excitation regions (1)-(3) in Fig. 6b. The emission image of region (1), selecting only the hole area, shows a single emission spot. Note that the elongation of the emission spot (horizontal direction in panel c-(1)) is due to the shape of the slit mask (see the details in SI). Since the emission-spot-decomposed CL map of the region (I) in panel d shows the field distribution of the dipole with the strong field along the *x*-axis edge of the hole, we understand that this center emission spot is dominantly formed by the localized mode of the hole. When the e-beam hits the flat silver area away from the hole, causing both transition radiation and emission of the local mode through SPPs, these two emission spots exhibit interesting interferences. Figures 6c-(2) and 6c-(3) show the emission images corresponding to the beam excitation on the flat film areas 0.5 µm away from the hole, as indicated by the (2) and (3) squares in panel b. As shown in Fig. 6c-(2), the emission image corresponding to excitation from the top side yields a single emission spot shifted to the upper side compared to the direct excitation of the hole in Fig. 6c-(1). In the corresponding emission-spot-decomposed CL photon map in Fig. 6d-(II), where only the emission from the upper part is selected, as indicated in Fig. 6c-(2), the contribution of transition radiation is clearly observed in the form of strong intensity over the flat silver film region. In contrast, the emission image from the lower part of the silver film (region (iii) in panel b) gives two separate emission spots, as shown in Fig. 6c-(3). The upper emission spot shares the same position as the direct nanohole radiation in panel c-(1), while the lower emission spot corresponds to the position of the transition radiation from the silver thin film. Indeed, in the emission-spot-decomposed CL photon map, using this lower spot (Fig. 6d-(III)), the bottom part of the flat silver film is brightened, which is opposite to the map of Fig. 6d-(II), constructed from the upper part of the emission spot. The difference in the spot shape depending on the upper (2) or lower (3) side excitations is related to the phase difference of the two emission sources, i.e. hole and TR, due to the asymmetry in the detection direction with respect to the *y-z* plane.[16] The relative phase of emission from the hole with respect to transition radiation from the e-beam position is different depending on whether SPPs reach the hole from the upper or lower sides of the film (i.e., from regions at different distances with respect to the mirror). Thus, the two emission spots appear to interfere constructively or destructively for lower or upper side excitation, respectively. Such interfering spot formation can be reproduced in the emission spot simulation assuming two separated emission sources corresponding to the hole and the transition radiation with different phases in SI.

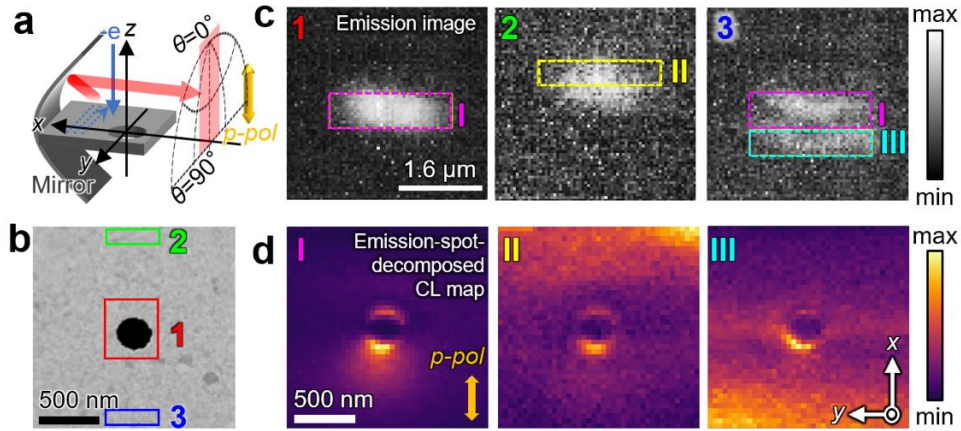

**Figure 6. Emission imaging of a single silver hole.** (a) Schematic illustration of a silver hole measurement with *p* polarization. The collected radiation is limited by the slit mask to the upward direction along the *x-z* plane, as indicated by a red stripe in the emission angle space. (b) STEM dark-field image of the probed silver nanohole with a diameter of 260 nm. (c) Emission images integrated over e-beam positions in the areas indicated by the rectangles 1-3 in panel (b). (d) Emission-spot-decomposed CL photon maps integrated over emission position space in the rectangles (I)–(III) in panel c. The direction of polarization is schematically shown by the yellow arrow in panel d-(I).

When two holes are adjacent within the attenuation distance of SPPs, the local modes of the two holes interact through the propagating SPPs, which emulates a system consisting of a receiver and a transmitter connected by a waveguide transporting the electromagnetic energy. We here demonstrate SEEM measurements of such a two hole-system, where a pair of silver holes are aligned along the $y$ direction (horizontal direction in the image), as shown in Fig. 7a. The holes are 500 nm in diameter and are separated by 2 μm, which should be long enough to avoid near-field interaction of local modes.[27] We choose to measure $s$-polarized emission to eliminate the contribution of transition radiation, and the same angular region as in the single hole measurements (Fig. 6) is selected. We take larger size holes to increase the hole-scattering signals relative to the single hole considered in Fig.6. Quadrupole components can be relevant for large holes, particularly at short wavelengths (see details of the influence of the quadrupole mode in SI). Under panchromatic imaging conditions, without wavelength selection in order to obtain a sufficient signal-to-noise ratio, we observe dominant dipole components similar to the smaller hole in Fig. 6: The wavelength-integrated emission image from the left hole (Fig. 7b-(1)) shows one emission spot corresponding to the local dipole mode excited at the left hole. The emission image from the right hole (Fig.7b-(3)) also has one spot, which is shifted along the $y$ direction from the left hole spot (Fig.7b-(1)). When the e-beam excites the area between the holes (region (ii) in Fig.7a), two emission spots are observed due to the excitation of both holes through SPPs, as shown in Fig.7b-(2). The emission-spot-decomposed CL map including both emission spots, corresponding to integration over e-beam positions within the area (I) in Fig.7b-(2), shows patterns that are consistent with electric field distributions of horizontally polarized dipoles of both holes, as shown in Fig. 7c-(I). In order to visualize energy transport from the left hole to the right one via SPPs, we extract the emission-spot-decomposed CL map (operation mode 2, see Fig. 1d) of the emission position region (II) in Fig. 7b-(3). As shown in the map in Fig. 7c-(II), the local mode of the right hole produces a dominant feature, while the dipole-like intensity distribution at the left hole is also visible. This means that emission from the right hole occurs upon left-hole excitation, directly showing evidence of signal transfer across holes mediated by SPPs. Thus, the SEEM method allows for a visual evaluation of energy transport between distant nanostructures.

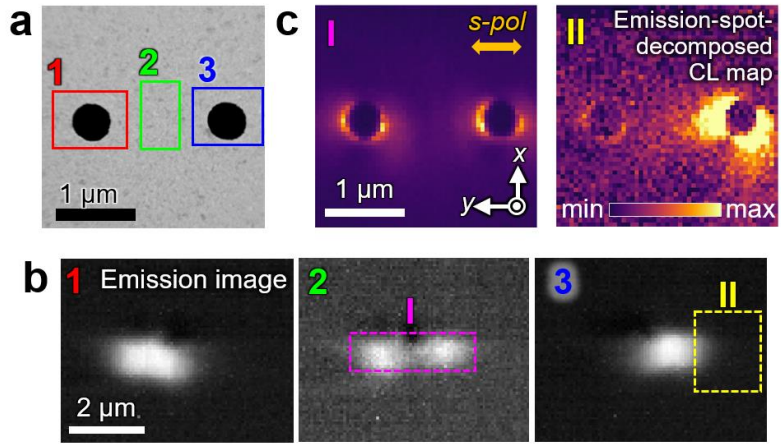

**Figure 7. Emission imaging of a pair of silver holes.** The detection conditions are the same as those of Fig. 6, but with s polarization emission (see double arrow in (c)) (a) STEM dark-field image of the probed silver holes with a diameter of 500 nm. (b) Emission images (operation mode (1), see Fig. 1c) integrated over e-beam positions in the areas indicated by the rectangles 1-3 in panel a. (c) Emission-spot-decomposed CL maps (operation mode (2), see Fig. 1d) integrated over the emission position space in the rectangles (I) and (II) in panel (b). The polarization direction *s* is schematically shown by the yellow arrow in the left panel.

## Conclusions

We have developed a new nano-optical imaging method to access the emission position space, which allows simultaneous 4D information acquisition including the 2D excitation mapping by e-beam and the 2D optical emission image corresponding to each excitation position. The projection plane of the 3D emission position space can be arbitrarily selected by the detection angle. By applying a center-of-mass analysis to the emission image, the resolution in the position of emission is pushed well below the diffraction limit of light. We have applied this SEEM-CL method to determine the spatial shift of the emission spot in a silver nanowire and revealed that the emission spot is modulated periodically as the e-beam is raster scanned and hits the nodes and antinodes of the plasmon modes in the nanowire. Using holes on a silver membrane, the spatial transport of the electromagnetic energy propagated by SPPs has been directly visualized by identifying both the excitation and the emission positions.

The introduced SEEM method constitutes a powerful approach for spatially analyzing the whole process of photon generation, from excitation to actual emission.[29] While we have presented application examples of coherent emission processes, the SEEM approach can also be applied to the evaluation of collective relaxation processes associated with incoherent emission, such as in semiconductors with carrier or exciton diffusion with a characteristic range from a few hundred nanometers to microns.[6, 30, 31] The effective spatial resolution in CL, which is limited by carrier diffusion, can be improved in the SEEM approach, for example, by selectively extracting the radiation emitted at the desired location to consider the effect of carrier diffusion. In consequence, the developed SEEM method enables a nanoscopic comprehensive analysis of excitation and emission that holds potential to unveil previously inaccessible information in optical materials and devices, opening a new era of nano-optical imaging.

## *Methods*

**Cathodoluminescence measurements**

A modified STEM (JEM-2100F, JEOL, Japan) with a Cs-corrector is used at an acceleration voltage of 80 kV.[27] The electron probe current is about 1 nA with a 20 mrad illumination half angle, which results in a 1 nm probe size. An aluminum parabolic mirror is mounted between the pole pieces of the objective lens, so that light emission from the sample at the focal point of the mirror is collected with a collection solid angle of more than 3 sr. The emission from the sample is collimated by the parabolic mirror, and guided out of the STEM column through the vacuum window. After traveling through a polarizer (*s* and *p* polarization), the emitted light is split into two ways by a beam splitter: to the camera for SEEM analysis (the optical path (I) in Fig. 1b); and to the spectrometer for angle-resolved spectroscopic CL (the optical path (II)). In the latter, the emission is focused on the spectrometer plane. The detection angle can be selected by a mask inserted on the emission angle planes of the both light paths (I,II). More details of the basic CL system are found in previous publications.[7, 17-19, 27] In the emission imaging system, the real-space emission position reflected by the beam splitter is focused on the camera plane. For a wavelength range selection in emission imaging, a bandpass filter is inserted in the optical path.

**Sample fabrication**

Silver nanowires (Silver nanowire A50SL (Ethanol dispersion)/ Filgen) are dispersed on a TEM grid with an elastic carbon supporting membrane. A nanowire with a diameter of 95 nm is selected observed in the experiment. Silver holes are prepared using colloidal lithography and film transfer,[32] where a free-standing carbon thin membrane with holes of diameters of 260 and 500 nm is fabricated and transferred to a TEM grid. Silver and $SiO_2$ films are deposited via sputtering on holey carbon films with thicknesses of 50 nm and 15 nm, respectively. Since one side of the hole film is covered with a carbon film, SPP propagation is limited to the silver / $SiO_2$ interface.

**Boundary-element method (BEM) simulation**

Simulations are carried out using an adapted version of the BEM that exploits the axial symmetry of the nanowire,[33] such that the sample boundary is one-dimensional and the number of parametrization points is radically reduced. The electric far-field amplitude associated with the CL emission produced by the e-beam excitation is separately calculated for each azimuthal angular number $m$, with a global dependence on azimuthal angle $\varphi$ encapsulated in an overall factor $e^{im\varphi}$. We achieve convergence using 700 parametrization points and $|m| \leq 4$. The COM analysis of the simulated field is performed by calculating the optical spot imaged through the parabolic mirror, as described in the Supplementary Information.

## *Contributions*

The manuscript was written through contributions from all authors. TM and TS conceived the research. TM, SO, and TS carried out the experiment. TM, FJGA and TS performed the analysis and calculation. All authors have given approval to the final version of the manuscript.

## *Corresponding author*

*sannomiya.t.aa@m.titech.ac.jp

## *Acknowledgements*

We thank Dr. Naoki Yamamoto for his valuable advice. The authors acknowledge the financial support from JSPS PD2(20J14821), JSPS Kakenhi (21H01782, 22H05033), JST FOREST (JPMJFR213J), the European Research Council (Advanced Grant No. 789104-eNANO), the Spanish MICINN (PID2020-112625 GB-I00 and Severo Ochoa CEX2019-000910-S), the Catalan CERCA Program, and Fundaciós Cellex and Mir-Puig.

## *Associated Content*

**Supporting Information** is available.

## *Ethics declarations*

**Competing interests**

The authors declare no competing interests.

## *References*

1. 1. Polman, A; Kociak, M.; García de Abajo, F.J. Electron-beam spectroscopy for nanophotonics. *Nature Materials* **2019**, 18, 1158-1171.
2. Kociak, M.; Stephan, O. Mapping plasmons at the nanometer scale in an electron microscope. *Chemical Society Reviews* **2014**, 43, 3865-3883.
3. Zamani, R.R.; Arbiol, J. Understanding semiconductor nanostructures via advanced electron microscopy and spectroscopy. *Nanotechnology* **2019**, 30, 262001.
4. Cherqui, C; Thakkar, N; Li, G.L; Camden, J.P.; Masiello, D.J. Characterizing Localized Surface Plasmons Using Electron Energy-Loss Spectroscopy. *Annual Review of Physical Chemistry* **2016**, 67, 331-357.
5. Kapetanakis, M.D. et al. Low-loss electron energy loss spectroscopy: An atomic-resolution complement to optical spectroscopies and application to graphene. *Physical Review B* **2015**, 92, 125147.
6. Edwards, P.R.; Martin, R.W. Cathodoluminescence nano-characterization of semiconductors. *Semiconductor Science and Technology* **2011, 26,** 064005.

7.  Yamamoto, N. Development of high-resolution cathodoluminescence system for STEM and application to plasmonic nanostructures. *Microscopy* **2016**, 65, 282-295.

8.  Krivanek, O.L. et al. Vibrational spectroscopy in the electron microscope. *Nature* **2014,** 514, 209-212.

9.  Saito, H. et al. Valley-Polarized Plasmonic Edge Mode Visualized in the Near-Infrared Spectral Range. *Nano Letters* **2021**, 21, 6556-6562.

10. Kumar, S.; Bozhevolnyi, S.I. Single photon emitters coupled to plasmonic waveguides: A review. *Advanced Quantum Technologies* **2021**, 4, 2100057.

11. Ben Nasr, F; Matoussi; A; Guermazi, S.; Fakhfakh, Z. Cathodoluminescence investigations of GaAs thin layers. *Proceedings of the Jmsm 2008 Conference* **2009**, 2, 827-833.

12. Coenen, T.; Haegel, N.M. Cathodoluminescence for the 21st century: Learning more from light. *Applied Physics Reviews* **2017**, 4, 031103.

13. Haegel, N.M. et al. Direct imaging of anisotropic minority-carrier diffusion in ordered GaInP. *Journal of Applied Physics* **2009**, 105, 023711.

14. Haegel, N.M. Integrating electron and near-field optics: dual vision for the nanoworld. *Nanophotonics* **2014**, 3, 75-89.

15. Baird, L. et al. Transport imaging for contact-free measurements of minority carrier diffusion in GaN, GaN/AlGaN, and GaN/InGaN core-shell nanowires. *Applied Physics Letters* **2011**, 98, 132104.

16. Sannomiya, T. et al. Cathodoluminescence Phase Extraction of the Coupling between Nanoparticles and Surface Plasmon Polaritons. *Nano Letters* **2020**, 20, 592-598.

17. Matsukata, T; Wadell, C; Matthaiakakis, N; Yamamoto, N.; Sannomiya, T. Selected Mode Mixing and Interference Visualized within a Single Optical Nanoantenna. *ACS Photonics* **2018**, 5, 4986-4992.

18. Matsukata, T. et al. Selection and Visualization of Degenerate Magnetic and Electric Multipoles up to Radial Higher Orders by Cathodoluminescence. *ACS Photonics* **2019**, 6, 2320-2326.

19. Matsukata, T; García  de Abajo, F.J.; Sannomiya, T. Chiral Light Emission from a Sphere Revealed by Nanoscale Relative-Phase Mapping. *ACS Nano* **2021**, 15, 2219-2228.

20. Betzig, E. et al. Imaging intracellular fluorescent proteins at nanometer resolution. *Science* **2006**, 313, 1642-1645.

21. Thollar, Z; Wadell, C; Matsukata, T; Yamamoto, N.; Sannomiya, T. Three-Dimensional Multipole Rotation in Spherical Silver  Nanoparticles Observed by Cathodoluminescence. *ACS Photomics* **2018**, 5, 2555–2560.

22. García  de Abajo, F.J. Optical excitations in electron microscopy. *Reviews of Modern Physics* **2010**, 82, 209-275.

23. Wei, H. et al. Plasmon Waveguiding in Nanowires. *Chemical Reviews* **2018**, 118, 2882-2926.

24. Pfeiffer, C.A; Economou, E.N.; Ngai, K.L. Surface polaritons in a circularly cylindrical interface - surface plasmons. *Physical Review B* **1974**, 10, 3038-3051.

25. Mkhitaryan, V. et al. Can Copper Nanostructures Sustain High-Quality Plasmons? *Nano Letters* **2021**, 21, 2444-2452.

26. Coenen, T.; Polman, A. Optical Properties of Single Plasmonic Holes Probed with Local Electron Beam Excitation. *ACS Nano* **2014**, 8, 7350-7358.

27. Sannomiya, T; Saito, H; Junesch, J.; Yamamoto, N. Coupling of plasmonic nanopore pairs: facing dipoles attract each other. *Light-Science & Applications* **2016**, 5, e16146.

28.     Schilder, N.J; Agrawal, H; Garnett, E.C.; Polman, A. Phase-Resolved Surface Plasmon Scattering Probed by Cathodoluminescence Holography. *ACS Photonics* **2020**, 7, 1476-1482.

29.     Meng, Y. et al. Optical meta-waveguides for integrated photonics and beyond. *Light-Science & Applications* **2021**, 10, 235.

30.     Hocker, M. et al. Determination of axial and lateral exciton diffusion length in GaN by electron energy dependent cathodoluminescence. *Journal of Applied Physics* **2016**, 120, 085703.

31.     Nagamune, Y; Watabe, H; Sogawa, F.; Arakawa, Y. One-dimensional exciton diffusion in GaAs quantum wires. *Applied Physics Letters* **1995**, 67, 1535-1537.

32.     Junesch, J.; Sannomiya, T. Ultrathin Suspended Nanopores with Surface Plasmon Resonance Fabricated by Combined Colloidal Lithography and Film Transfer. *ACS Applied Materials & Interfaces* **2014**, 6, 6322-6331.

33.     García de Abajo, F.J.; Howie, A. Retarded field calculation of electron energy loss in inhomogeneous dielectrics. *Physical Review B* **2002**, 65, 115418.

# Supporting Information for

# "Simultaneous nanoscale excitation and emission mapping by cathodoluminescence"

*Taeko Matsukata[1], Shintaro Ogura[1], F. Javier García de Abajo[2,3], Takumi Sannomiya[1]*

AUTHOR ADDRESS

[1] Department of Materials Science and Technology, Tokyo Institute of Technology, 4259 Nagatsuta Midoriku, Yokohama 226-8503, Japan

[2] ICFO-Institut de Ciencies Fotoniques, The Barcelona Institute of Science and Technology, 08860 Castelldefels (Barcelona), Spain

[3] ICREA-Institució Catalana de Recerca i Estudis Avancats, Passeig Lluís Companys 23, 08010 Barcelona, Spain

## S1. Details of the measurement procedure

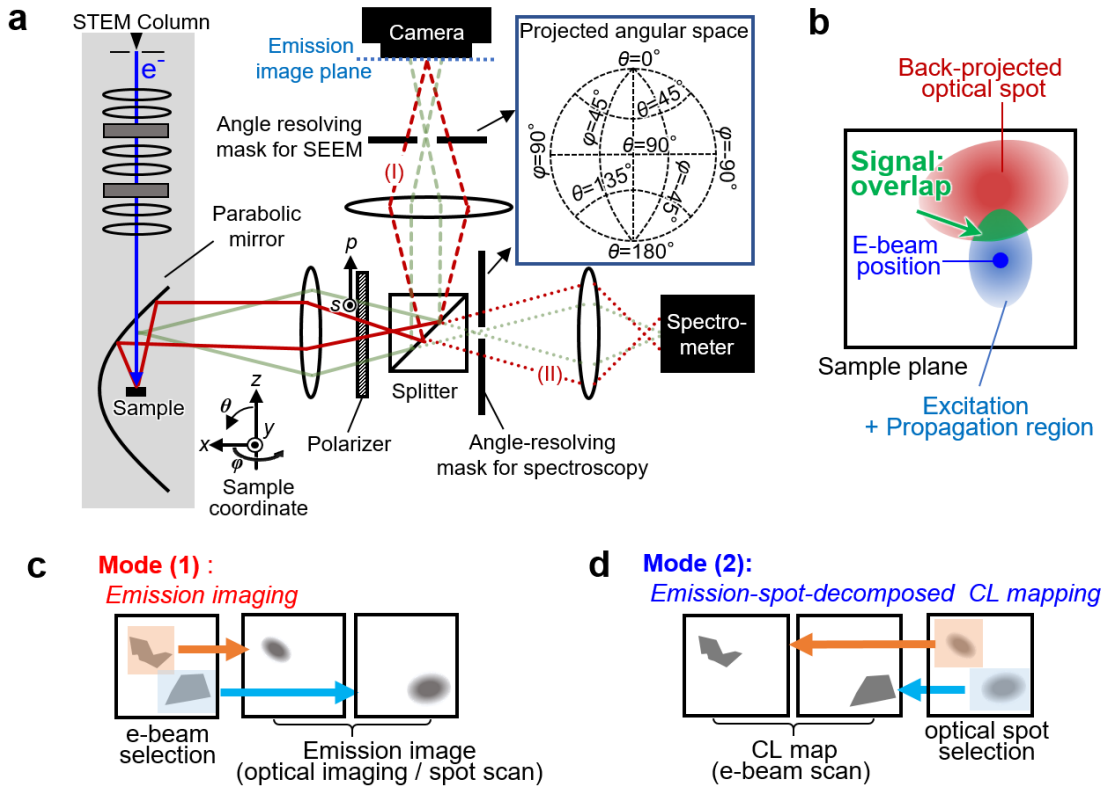

**Figure S1. Detailed diagram of the measurement setup**. (a) Optical ray path in the setup. The CL signal is detected (I) by a camera for emission position imaging and (II) by an spectrometer for CL spectroscopy. Ray paths associated with emission-space and reciprocal-angular-space imaging are drawn as red and green lines, respectively. The emission angle can be selected in the projected angular space with the use of an angle-resolving pinhole mask, as shown in the diagram. We use identical selection ranges for emission imaging and spectroscopy. (b) Illustration to describe the signal generation, which is the overlap of the back-projected optical spot and the e-beam excitation area including optical mode propagation in the specimen. (c,d) Schematics of the image data extraction from the obtained 4D dataset. (c) Operation mode (1): optical emission images are extracted from selected fixed e-beam excitation positions. (d) Operation mode (2): e-beam scan CL images are extracted from selected fixed optical spot areas.

## S2. Projection of the emission position space without angle selection

Without inserting the angle-selecting mask, the optical axis for emission imaging is parallel to the $x$ direction and the mirror spans an acceptance solid angle of ~$3\pi$ sr. This section describes the response of the emission-spot shift with respect to the shift of the emitter in the sample real space. A 2$\mu$m ZnS:Pb particle is used as the emitter sample and moved by $\pm 2$ $\mu$m in the $x$, $y$, or $z$ directions, as schematically shown in Figs. S2a, S2d, and S2g, respectively. For emission imaging, the CL signal is integrated while the electron beam is scanned over the entire scan area of the STEM image. In the STEM darkfield image, which projects the $x$-$y$ plane of the sample space, a shift of the sample along $x$ is observed as an in-plane position shift, as shown in Fig. S1b. In contrast, such sample shift along the $x$ direction causes a defocus of the emission spot in the corresponding emission image (Fig. S2c). However, a sample shift along the $y$ direction is observed as an in-plane position shift (horizontal direction) in the emission image (Fig. S2d-f). Also, a shift of the sample along $z$, corresponding to the defocus in the STEM image, is instead observed as an emission spot shift along the vertical direction in the emission image (Fig. S2i). Thus, using the entire parabolic mirror without an angle-selecting mask, the projection plane of the emission image corresponds to the $y$-$z$ plane, which does not match the STEM $x$-$y$ image projection plane. Since the emission position image in this configuration provides information on the $z$ direction, 3D emission imaging is potentially possible.

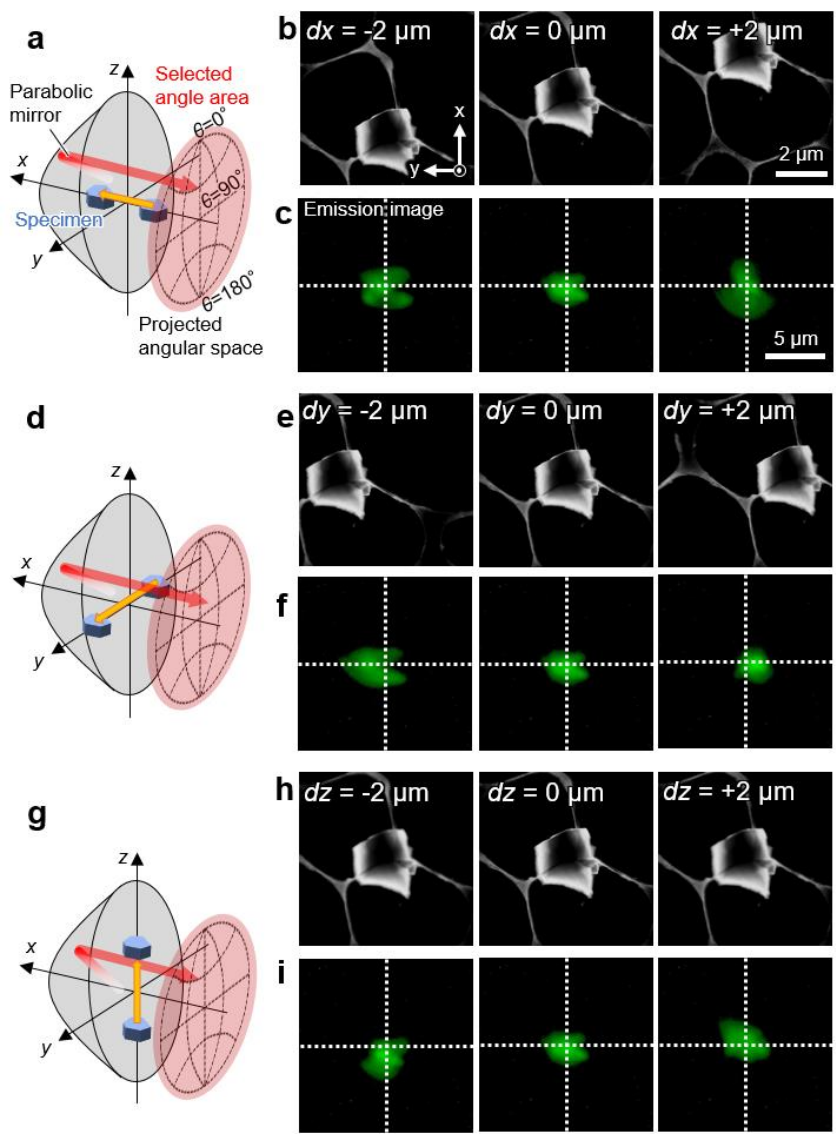

**Figure S2. Changes in the emission images produced by sample shifts without an angle-resolving mask.** (a, d, g) Schematic illustrations of sample shifts along the (a) *x*, (d) *y*, and (g) *z* directions. The collected angle range (corresponding to the size of the mirror) is indicated by a red hatching in the emission angle space. (b, e, h) STEM dark-field images of the probed ZnS:Pb particle with a diameter of around 2 µm. The sample is moved from the center by ±2 µm (left and right columns) along the (b) *x*, (e) *y*, and (h) *z* directions. Since the imaging plane of the STEM dark-field image is on the *x-y* plane of the sample space, the *z*-shift of the sample corresponds to defocusing of the STEM image. (c, f, i) Emission images obtained while scanning the electron

beam over the whole imaging area of the STEM images. The emission image corresponds to the projection of the *y-z* plane, and thus, the sample shift along *x* causes a defocus of the emission image.

## S3. Excitation-position-decomposed emission image

The excitation position distribution that contributes to a certain emission spot of Fig. 3 in the main text is extracted from the 4D data set (2D excitation and 2D emission). In this section, in contrast to Fig. 3, the emission image corresponding to a specific excitation position region is extracted from the same 4D dataset. The STEM image in Fig. S3a is the same as Fig. 3b in the main text, and the CL map of Fig. S3b is the same image as Fig. 3d-(I), which consists of the signals integrated over the entire emission image. Figure S3c shows the colored superposition of the emission images corresponding to the selected particles as indicated by color-matching boxes in Fig. S3b. Thus, two modes of post-processing operation are possible in the presented SEEM method: (1) emission position imaging: optical spot scanning while integrating over a selected excitation (e-beam) positions (Fig. S3c); and (2) e-beam scan while integrating over emission positions (Fig. 3d in the main text).

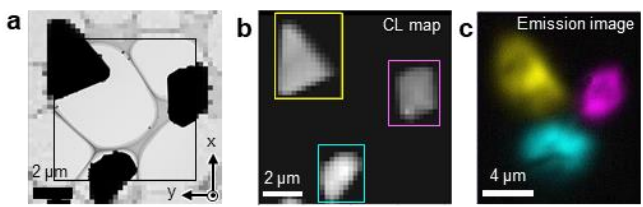

**Figure S3. Emission image resolved on excitation position.** (a) STEM bright-field image of the same probed ZnS:Pb particles as in Fig. 3 of the main text, supported on a microgrid. (b) CL map integrated over the entire emission image (mode 2). (c) Colored superposition of three emission

images (mode 1) obtained by integrating the signal over e-beam positions within the areas indicated by the color-coordinated boxes in panel (b).

## S4. Calculation of the emission image on the camera

We here calculate the shift and shape of the emission spot on the emission image under paraxial conditions. The focal length of the parabolic mirror is set to 1.5 mm and the light wavelength to 500 nm to emulate the experiment. In the remainder of this discussion, we assume these conditions unless otherwise stated. As a first approximation, the position shift on the sample plane induces a phase shift $\Delta$ due to the difference in the optical path, which is described in Fig. S4a. This phase difference can be expressed as $\Delta = 2\pi (R'-R)/\lambda$, where $\lambda$ is the wavelength of the emitted light and primed quantities refer to the shifted sample. The distance $R$ from the origin (focal position of the parabolic mirror) to the mirror position $\mathbf{R}(Y, Z)$ is a function of the emission beam position on the aperture plane space. Similarly, we have $\mathbf{R}'(Y, Z)$ for the shifted sample. The wave function on the aperture plane is a product of the wave $A$ from the sample and the pupil function $A_p$ of the aperture: $\Psi = A_p \times A$, where $A = A_0 \exp(-i\Delta)$ and $A_0$ are the complex amplitudes spread over the detection solid angle. The spot intensity $I(x', y')$ on the camera plane can be calculated by Fourier transforming the wave function on the aperture plane (i.e., $I = \left| \text{FT}[\Psi] \right|^2$).

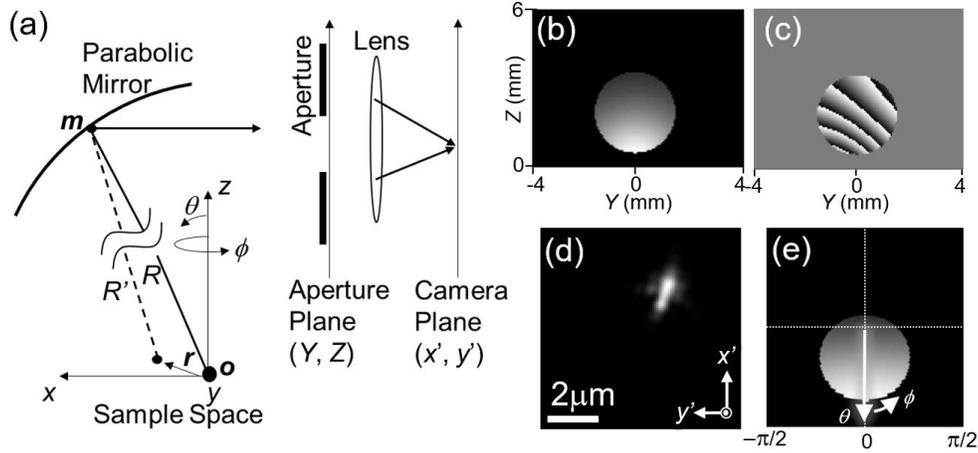

**Figure S4. Calculation scheme for the emission spot image.** A parabolic mirror with a focal length of 1.5 mm and a wavelength of 500 nm are assumed. (a) Geometry for the emission image calculation. (b) Intensity and (c) phase of the wave function on the aperture plane. (d) Calculated beam on the camera plane. A non-polarized spherical wave from a single emitter is assumed. (e) Wave intensity projected on the $\theta$-$\varphi$ space. An aperture diameter of 3 mm is placed at $(Y,Z) = (0$ mm, 2 mm). The sample position is set to $(x, y, z) = (1$ μm, -1 μm, 1 μm). The scale bar corresponds to the size on the sample plane.

## S5. Simulated projection-plane selection of the emission position

The optical axis of the imaging system can be selected by appropriately choosing the detection angle. In this section, the response of the emission image to the shift of the source depending on the detection angle is clarified by calculating the emission image. We assume an unpolarized spherical wave source with a wavelength of 500 nm located near the focal position of the parabolic mirror and with the radiation angle selected by a circular aperture with a diameter of 3 mm. Fig. S5a shows the intensity distribution in the radiation angle space projected on the aperture plane with an angle-selecting aperture placed at $\theta = 0$ ° (i.e., the same conditions as in Fig. 2 of the main text). Figure S5b shows the intensity distribution mapped onto polar coordinates. Figure S5c shows the emission images obtained with the source shifted along the *x*, *y*, and *z* directions. Since the optical axis for imaging is now approximately lying along the *z* axis, an in-plane shift of the source on the *x*-*y* plane is projected as an in-plane shift on the emission image plane *x'*-*y'*. The *z*-shift of the source causes a defocus in the emission image.

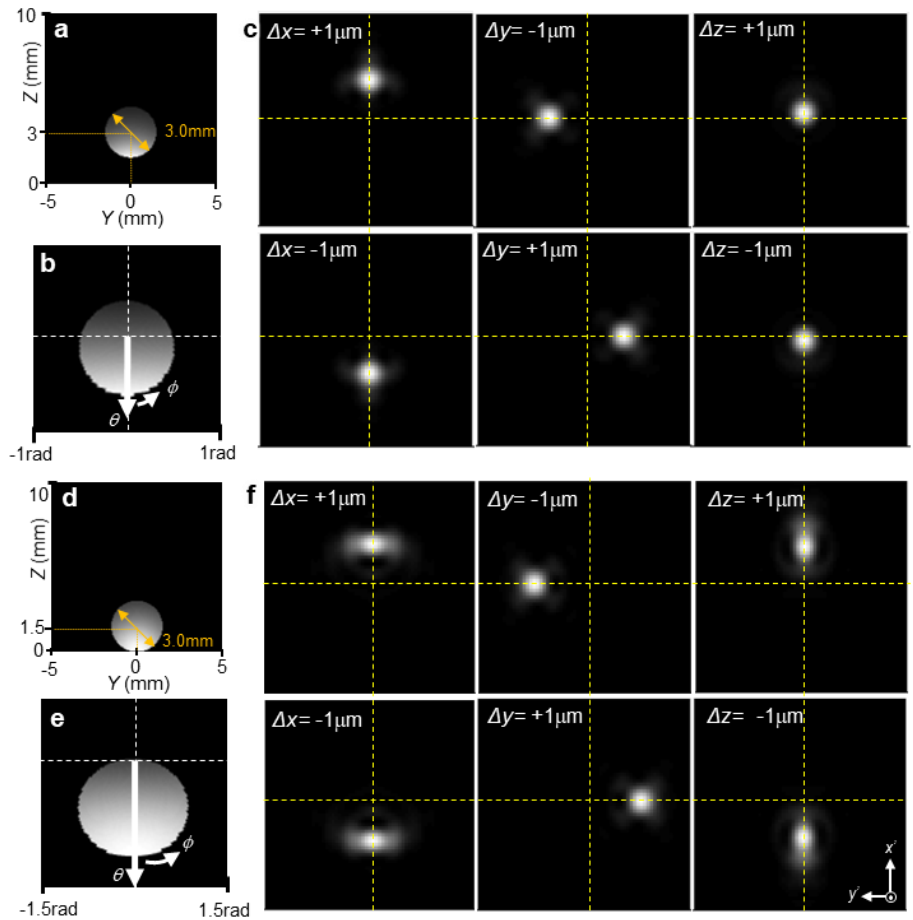

**Figure S5. Calculated emission image of an unpolarized spherical wave source.** An aperture with a diameter of 3 mm is placed at $(Y, Z) = (0 \text{ mm}, 3 \text{ mm})$ for panels a-c and $(Y, Z) = (0 \text{ mm}, 1.5 \text{ mm})$ for panels d-f. The emission intensity distributions in angle space are displayed (a, d) on the aperture plane and (b, e) on the angular plane of the specimen in polar coordinates. (c, f) Emission images with source shifts. The displacement distance and direction are labeled in each panel.

When the selection angle $\theta$ is made larger, a shift in the emission spot along $z$ also causes a corresponding shift in the emission image, as shown in Fig. S5d-f, which emulates the configuration considered in Fig. 4 of the main text. Figure S5f shows the calculated emission images with shifts of the source along $x$, $y$, and $z$. A source shift along $y$ still corresponds to a pure $y'$ shift of the emission image, while the $x$ and $z$ shifts of the source are mixed on the image plane along the $x'$ axis.

## S6. Shape of the emission spot produced by sample position shifts

In the experiment, the sample position is adjusted such that the electron beam position matches the focal position of the parabolic mirror. The imaged emission spot shape changes slightly as the emitter is shifted from the focal position of the parabolic mirror. Here, we calculate the emission images when the emission point is shifted by ± 1 μm along the $x$, $y$, and $z$ axis directions from the focal position. The emission source is assumed to be an unpolarized spherical wave. The calculated emission images for emission point displacements of $x$, $y$ = ± 1 μm from the origin are shown in Fig. S6. The $z$ position is fixed at (a) $z = 0$ μm and (b) $z = 1$ μm. Without displacement from the focal position (Fig. S6a), the emission spot is nicely round, while with a displacement within the $x$-$y$ plane, the shape of the emission spot is elongated with some blurring, which is more noticeable when introducing a displacement along $z$ (Fig. S6b).

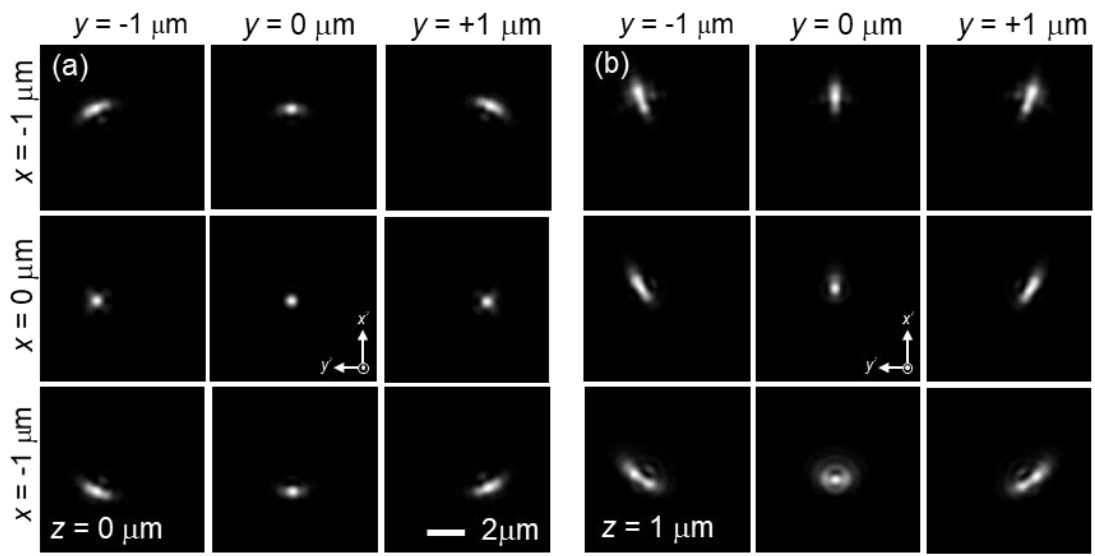

**Figure S6. Calculated emission images at different sample (emitter) positions assuming non-polarized spherical waves.** An aperture of 3 mm diameter is placed at $(Y, Z) = (0 \text{ mm}, 2 \text{ mm})$ and images are calculated for (a) a sample position $z = 0$ μm and (b) a sample position $z = 1$ μm. The scale bar refers to the size on the sample plane.

## S7. CL spectrum of the single hole with a diameter of 260 nm

A CL spectrum of a single 260 nm hole as in the main text (Fig.6) is shown in Fig. S7.

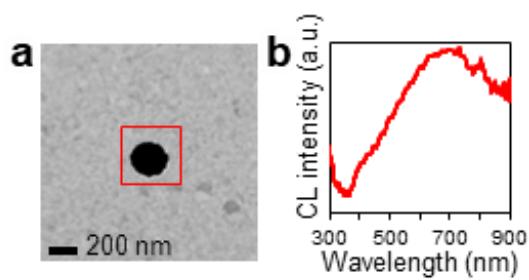

**Figure S7. CL spectrum of the single silver hole measured in Fig. 6 of the main text.** (a) STEM dark-field image of the hole (260 nm diameter). (b) CL spectrum integrated over the area of the red box in (a).

**S8. Emission spot shape with a slit mask**

In Fig. 6c of the main text, the emission spot on the emission image is elongated along the *y* direction due to the shape of the slit mask used for emission-angle selection. Here, we clarify the effect of the shape of the slit mask on the emission spot shape by calculating the emission image. We assume a spherical wave from the focal position of a parabolic mirror with a focal length of 1.5 mm, emulating the experiment. We also account for the hole along the electron-beam path on the parabolic mirror to better reproduce the emission image. Figures S8a and S8b show the emission intensity distributions projected onto the aperture plane, or the *Y-Z* plane, with angle selection by slits of 4.0 mm × 1.5 mm (Fig. S8a) and 4.0 mm × 1.0 mm (Fig. S8b). The calculated emission spots are shown in panels c and d, respectively. The wider slit (Fig. S8a) yields a more focused spot (Fig. S8c) compared with the spot of the narrower slit (Fig. S8b,d).

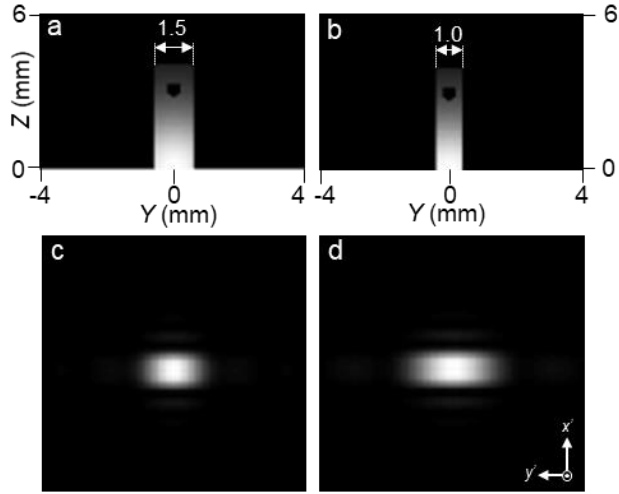

**Figure S8. Calculated emission spots with slit masks of two different widths.** (a,b) Emission intensity distributions projected onto the *Y-Z* aperture plane. The slit-shaped apertures select emission-angle areas of (a)$Y = -0.75$ to $0.75$ mm and $Z = 0$ to $4$ mm, and (b) $Y = -0.5$ to $0.5$ mm and $Z = 0$ to $4$ mm. (c,d) Calculated emission spots corresponding to the slit apertures of panels (a) and (b), respectively. An unpolarized spherical wave source is assumed.

## S9.  Interfering emission spots from transition radiation and hole

Figures 6c in the main text shows the emission spots obtained with constructive or destructive interference of the transition radiation emanating from the electron beam position and radiation from local modes of the hole. Here, we calculate the interfering emission spots of two spatially separated emitters. The angle selection is the same as that in the experimental setting. The hole situated at the origin is modeled as the superposition of an electric dipole polarized along the $z$ axis and a magnetic dipole polarized along the $y$ axis.[1,2]  The transition radiation is modeled as an electric dipole polarized along the $z$ direction, which is located at $(x, y, z) = (+ 0.5$ µm, $0$ µm, $0$ µm), as shown in the schematic illustration of Fig. S9a. Figures S9b and S9c show the emission spots associated with the hole and transition radiation. With a phase difference $\delta = \pi/2$ between the two emitters, the two emission spots are discernable in the emission image (Fig. S9d), while for a phase difference $\delta = -\pi/2$, only a single overlapping emission spot is observed (Fig. S9e). This is consistent with the experimental results shown in Fig.6 in the main text.

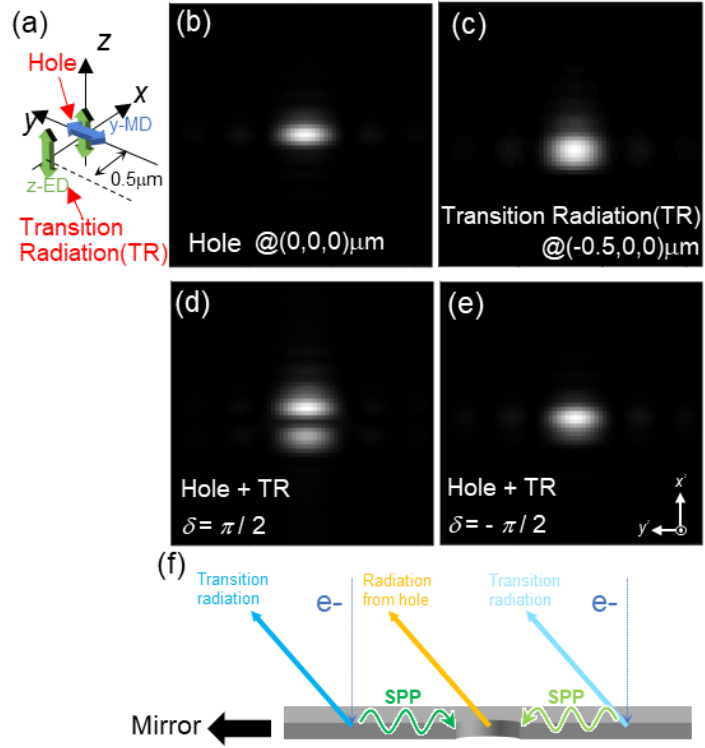

**Figure S9. Calculated emission images for two emitters, corresponding to transition radiation and emission from a hole, which are separated by a 0.5 μm distance along the *x*-direction.** The transition radiation is represented by an electric dipole oriented along the *z* direction and the hole emission by a combination of a *z* electric dipole and a *y* magnetic dipole.[1,2] A rectangular slit mask of 1.5 mm × 4.0 mm is applied. This simulation describes the emission from the hole (*z* electric dipole + *y* magnetic dipole) and the transition radiation (*z* electric dipole) observed in Fig. 6 in the main text. The images are magnified by a factor of two compared to Fig S6. (b,c) Spot images of the individual emitters. (d,e) Spot images of two emitters with different relative phases. (f) Schematic illustration of e-beam excitation of TR, SPPs, and the hole mode. The phase of the magnetic hole mode is flipped depending on the SPP launching direction.

**S10. Interference of electric dipole and quadrupole**

In Fig. S10, we show experimental results for a 500 nm silver hole showing that the contribution of the electric quadrupole produces two intensity maxima in the emission image. This effect is clearly observed in the wavelength range of 400-500 nm, while at in longer wavelength range (600-700 nm) the a single spot is observed due to the dipole nature of the excitation (Fig.S10c). Under panchromatic imaging conditions, as shown in the main text (Fig. 7), the spot from a single 500 nm hole gives rise to a single intensity maximum because of the dominant contribution of the dipole, which is also clear in the spectrum in Fig. S10b. Such splitting can be reproduced by calculating the *s*-polarization-filtered emission spot with a quadrupolar component. We introduce a magnetic dipole and a magnetic quadrupole placed at the origin to mimic the radiation field produced by a hole in a perfect electric conductor,[1,2] as shown in the schematic diagram of Fig. S11a. We do not include a perpendicular electric dipole because it contributes negligibly to the *s*-polarized emission. The magnetic dipole is polarized along the *x* direction and the magnetic quadrupole lies on the *x-y* plane and is placed as shown in the scheme. Both of these contribute strongly to the *s*-polarized light detection under the experimental conditions. Figures S11b, S11c, and S11d show emission images of the magnetic dipole, the magnetic quadrupole, and the superposition of both. In the contribution of the quadrupole component, the emission spot splits into two, just as in the experiment reported in Fig. S10c.

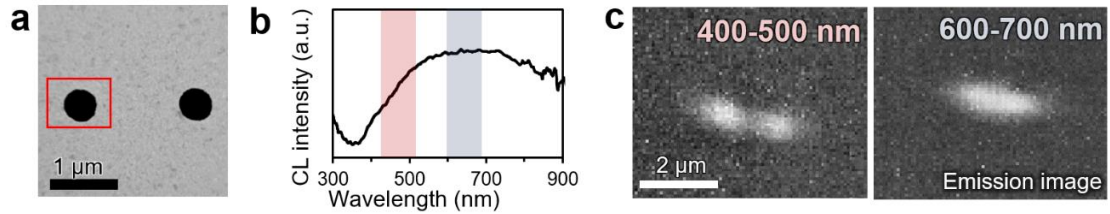

**Figure S10. Influence of the quadrupolar component.** (a) STEM bright field image highlighting the studied 500 nm hole (on the left). (b) Integrated CL spectrum in the area indicated in panel a. The detection conditions are the same as those of Fig. 6 in the main text, but with s polarization. (c) Emission images of the left hole integrated over the area highlighted in panel (a) within the wavelength ranges of 400-500 nm (left) and 600-700 nm (right).

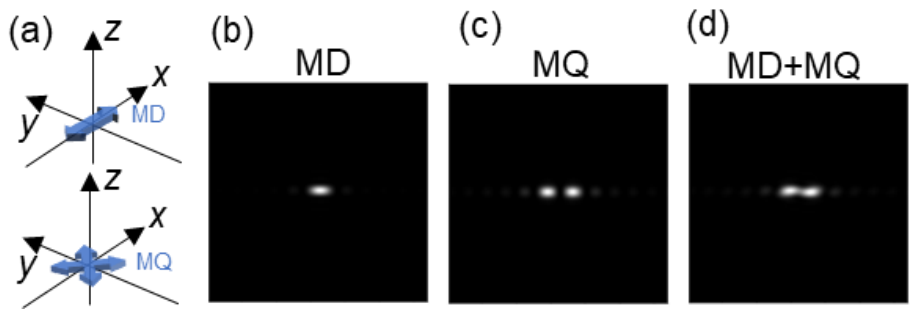

**Figure S11. Dipolar and quadrupolar spot profiles.** *s*-polarized emission images calculated for an emitter consisting of a magnetic dipole and a quadrupole. A slit mask of 1.5 mm x 4.0 mm is assumed as in the experiment of Fig.7 in the main text. The contribution to the emission spot from the quadrupole component splits into two.

## *References*

1       Rotenberg, N. *et al.* Magnetic and electric response of single subwavelength holes. *Physical Review B* **2013**, 88, 241408(R).
2       Bethe, H. A. Theory of diffraction by small holes. *Physical Review* **1944**, 66, 163-182.